\documentclass[twocolumn]{openjournal}
\usepackage{multirow}
\usepackage{hyperref}

\begin{document}

\title{Evidencing the interaction between science fiction enthusiasm and career aspirations in the UK astronomy community}
\author{Elizabeth~R.~Stanway}
\affiliation{Centre for Exoplanets and Habitability, University of Warwick, Gibbet Hill Road, Coventry, CV4 7AL, UK.}
\email{email: e.r.stanway@warwick.ac.uk}

\begin{abstract}
\noindent The anecdotal connection between an interest in science fiction and career aspirations in astrophysics is well established. However strong statistical evidence for such a connection, and a quantitative assessment of its prevalence, has been missing. Here I report the results of two surveys examining the connection between science fiction enthusiasm and astronomical careers - first a case study of the University of Warwick Astronomy and Astrophysics group, carried out in February 2021, and second a larger survey of attendees at the UK National Astronomy Meeting in July 2022. In both surveys, a significant majority of respondents expressed an interest in science fiction. In the larger survey, 93\% of UK astronomers (223 of 239 respondents) expressed an interest in science fiction, while 69\% (164) stated that it had influenced their life or career choices. This study provides strong statistical evidence for the role of science fiction in influencing the adoption of astronomical careers.
\end{abstract}

\section{Introduction}\label{sec:intro}

As an extensive body of literature discusses, an interest in or study of science fiction can shed a unique light onto aspects of science and society, and can also act to modify perceptions of science and its applications \citep[see e.g.][and citations therein]{doi:10.1177/2158244017723690}. Anecdotally, the evidence for a connection between an interest in physical sciences and an interest in science fiction is strong. Well-known individuals in a range of scientific and technical fields have spoken or written publicly about their interest in SF. The connection between SF and astronomy or space science is perhaps even stronger, since both typically involve the envisaging and exploration of aspects of the universe  
which lie outside our everyday experience. 

Examples of high-profile, self-professed SF fans from the astronomical community include planetary scientists Colin Pillinger and John Zarnecki, cosmologist Stephen Hawking 
and astrophysicist Carl Sagan, while discussions of SF, SF tropes and in-jokes are also a common feature amidst communities of scientists, including on social media. These even occur in the peer-reviewed academic literature. Examples of astronomical software, surveys and projects which reference science fiction in their names are presented in table \ref{tab:sfexamples}. The examples listed represent only a small fraction of the science fiction references embedded within contemporary professional astrophysics.

\begin{table*}
 \caption{Examples of science fiction references used in astronomical surveys and software.}\label{tab:sfexamples}

\begin{tabular}{p{0.4\columnwidth}lp{1.3\columnwidth}}
Title & Origin & Description\\
\hline\hline
BATMAN & & \multirow{3}{1.3\columnwidth}{Exoplanet lightcurve and parameter fitting codes, each named after the eponymous SF franchises  \citep{2015PASP..127.1161K,2018MNRAS.477.2613L,2019ESS.....432643L}.}\\
SPIDERMAN & &\\
TERMINATOR & &\\
MINBAR & Babylon 5 & the Multi-INstrument Burst ARchive, a repository for data on stellar explosions, named after a race from Babylon 5 \citep[][Galloway priv. comm.]{2020ApJS..249...32G}.\\
TARDIS & Doctor Who & supernova modelling code. rapid spectral modelling, open source software in which the origin of the name in Doctor Who is never explicitly stated \citep{2014MNRAS.440..387K}.\\
Cardassian Expansion & Star Trek &  A cosmological model first named in a paper which states explicitly that `The name Cardassian refers to a humanoid race in Star Trek whose goal is to take over the universe, i.e. accelerated expansion. This race looks foreign to us and yet is made entirely of matter.' \citep{2002PhLB..540....1F}\\
BoRG & Star Trek & the Brightest of Reionization epoch Galaxies survey, named for the aliens in Star Trek \citep{2011ApJ...727L..39T,2016ApJ...817..120C}.\\
DS9 & Star Trek & Image viewing software. Full name SAOImage DS9, the successor to SAOImage and SAOImage TNG \citep{2003ASPC..295..489J}.\\
Tatooines & Star Wars & used to describe circumbinary planets, as in the TATOOINE (The Attempt To Observe Outer-planets In Non-single-stellar Environments) radial velocity survey programme \citep{2010ApJ...719.1293K}, and in publicity surrounding Kepler-16b \citep{2019NewAR..8401515D}.\\
ACBAR & Star Wars & the Arcminute Cosmology Bolometer Array Receiver. In reference to the Star Wars character. It’s a trap! (A photon trap) \citep{2003ApJS..149..265R}.\\
`A long time ago in a galaxy far, far away' & Star Wars & Catchphrase frequently used e.g. as a paper title when discussing galaxies in the distant Universe \citep{2022arXiv220712474F} or in press releases.\\

\end{tabular}
\end{table*}

The impact of science fiction on perceptions of science is also recognised in areas which include public engagement and science education \citep[e.g.][]{2015CSSE...10..921H,2021IJSEd..43.2501D}. In particular, the field of exoplanet and habitability research has found its deep roots in fictional imaginaries of benefit for framing public science discourses \citep[e.g.][]{2016AsBio..16..325B,2016ApJ...826..225M,2019arXiv191000940S,eldridge2021s,stanway2021changing,stanway2021s,2021EPSC...15..870N}. The NASA Mission Concept Report for the proposed Habitable Exoplanet Observatory \citep[HabEx, ][]{2018arXiv180909674G}, for example, even notes that five of their nine primary targets have a popular culture connection - not because this influenced their selection, but because it is relevant to the degree of popular support for and public engagement with an expensive  space telescope. 

However the impact of science fiction on the professional community has been largely neglected in the literature, with the majority of studies relying on indirect evidence or small samples. \citet{2017SpPol..41...36A}, for example, discussed the link between science, philosophy, policy, and public perception as related to the {\em Star Trek} franchise. This work explored both the scientific content of the television series and its positioning in the context of social, political and technological assumptions regarding space utilisation. They note that `Star Trek made science fascinating. The characters were working as a solid and complex team to lead scientific missions while helping other people' and that `The consequence is the open desire for many fans to make it real. Fans who thus became scientists'. Such statements are based largely on anecdotes, including the assertion that the original series of {\em Star Trek} (1966-1969) helped to influence a doubling in the intake of female science and engineering students between 1970 and 1974 \citep{10.2307/1576348}. 

A less anecdotal and more statistic analysis of the interaction between science fiction and career choices in the sciences was undertaken by \citet{orthia_2019}. They carried out a survey of respondents to a Facebook post distributed by the Australia-based {\em ScienceAlert} news feed, seeking comments from those interested both in science and in the long-running science fiction television series {\em Doctor Who} (BBC TV, 1963-1989, 2005-present). Of 575 individuals sampled, 69\% had at least a Bachelor's-level education in a science topic, although only 35\% currently worked in a science-oriented job. Most of those surveyed felt that the science fiction had influenced their world-views and perception of science. However, of those who responded, between 75 and 78\% felt that their interest in {\em Doctor Who} had no impact on their career or education choices. While this still suggested a non-negligible fraction had been or may have been so influenced, the broad scope of `science' in the question posted, together with the specific framing of the survey to address Doctor Who enthusiasts, limits its interpretation in the context of the astronomical community. 

Crucially, by surveying only those already engaged with the genre, previous work has constrained the scope of possible responses from those who do not enjoy science fiction. Previous studies have also either focussed on the fan-base of a particular science fiction franchise, or those engaging with science fiction-related social media streams \citep[e.g.][]{doi:10.1177/2158244018780946}, rather than addressing professional scientists directly. To date, there has not been clear statistical evidence published for the strength of science fiction enthusiasm amongst the astronomical community.

This article presents two studies exploring engagement with science fiction amongst UK astronomers. In section \ref{sec:warwick}, I present a survey of an individual university astrophysics group, and discuss its responses. In section \ref{sec:NAM}, I present a larger statistical sample of data gathered at the UK National Astronomy Meeting 2022.  In section \ref{sec:disc}, I consider the implications of these survey results in a broader context. Finally, in section \ref{sec:conc}, I summarise my main conclusions.

\section{Warwick Astronomy \& Astrophysics: A case study}\label{sec:warwick}

\subsection{Methodology}
 Interest in SF amongst the Astronomy and Astrophysics research cluster at the University of Warwick was surveyed\footnote{Note: this survey was originally reported informally in a blog entry which can be found at \url{www.warwick.ac.uk/CosmicStories/astronomers_and_science}.} in February 2021. Since the advent of the Covid-19 pandemic, the group (a subset of the university's Physics department) had used the networking service {\it Slack} for both social and academic interactions. Similar to a discussion forum environment, but restricted to members of a specific organisation or project, this platform has permitted a relaxed atmosphere which encourages group members at all career stages to interact (although inevitably individual degree of engagement varies). The following question was posted on the \#general channel, which was designated for casual conversation:\\
\\
 Q: Are you a fan of, or have you been inspired by any of the following Science Fiction? (please click all that apply, even if your answer is no, I'd love to sample most of the group).
\\
1. Star Trek,\\
2. Star Wars,\\
3. Doctor Who,\\
4. Other SF (feel free to specify in comment),\\
5. Not a science fiction fan.\\

Responses were collected using the {\it Polly} survey application within {\it Slack}. This permits multiple responses to be recorded for each individual. A number of free-form text comments were also posted in response to the survey. The full response from any individual could be identified by the survey organiser, but not by other participants. 

\subsection{Results}

\begin{figure}
    \centering
    \includegraphics[width=\columnwidth]{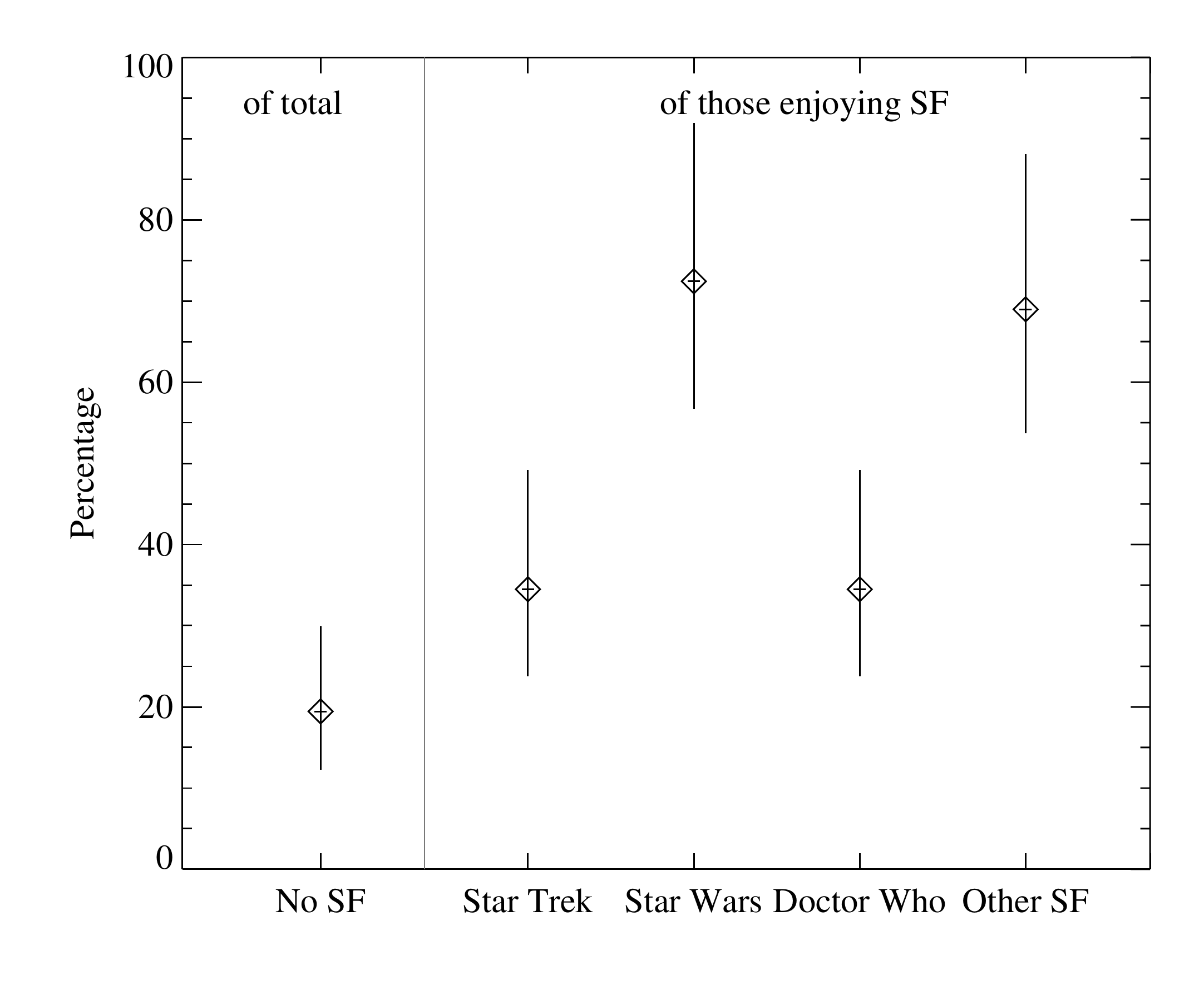}
    \caption{The results of a survey of SF interest amongst members of the Warwick Astronomy \& Astrophysics group in February 2021 (N=36). Error bars indicate small number sampling uncertainties \citep{1986ApJ...303..336G}.}
    \label{fig:warwick}
\end{figure}

Responses were collected from a total of 36 individuals, out of about 45 regular users of the forum at the time (ranging from new graduates to senior professors). The results are presented in figure \ref{fig:warwick}. Only seven of the 36 respondents (19\%) did not consider themselves a fan of or inspired by science fiction. Of the remaining four-fifths (N=29) of the group, 10 (34\%) selected {\em Star Trek}, 21 (72\%) selected {\em Star Wars}, 10 (34\%) selected {\em Doctor Who}, and 20 (70\%) selected other science fiction. 

Examples of influential SF mentioned in comments included novels such as Isaac Asimov's {\em Robots} and {\em Foundation} series (1942-1003), Arthur C.~Clarke's {\em Rendezvous with Rama} (1973), Dan Simmons’ {\em Hyperion Cantos} (1989-1997), Iain Banks {\em Culture} series (1987-2012) and Stanislaw Lem’s {\em Solaris} (1961). Also mentioned were comics including {\em Val\'erian and Laureline} (1967-2010),  television series including {\em Battlestar Galactica} (1978-9, 2003-9), {\em The X-Files} (1993-2002) and {\em Futurama} (1993-2003), and films including {\em 2001: A Space Odyssey} (1968).

Several participants observed that, while they were enthusiastic about science fiction, real world news stories (such as the Apollo programme moon landings or shuttle missions) or factual reporting (such as the long-running BBC television programme {\em Sky at Night} (1957-present) or Carl Sagan's landmark television documentary series {\em Cosmos}, first broadcast in 1980) had more impact on them than fiction. Perhaps unsurprisingly, comments to this effect were more common from more senior members of the group, although not restricted to them.

Other narrative comments indicated that an interest in science had preceded an interest in science fiction in a number of cases, with a typical example stating that `Sci-Fi plays up to the sense of wonder and awe that I got from reading my books about the planets as a five year old'. One participant also made the point that their interest in science fiction exists `in spite of' their enjoyment of science, rather than because of it, with their rational brain disengaged and no attempt made to watch critically. 

These responses confirmed the strong interest in science fiction amongst professional astronomers but also highlighted the need to consider enthusiasm for and inspiration by science fiction as separate axes of diversity amongst the professional astronomy community.

\section{Surveying the UK National Astronomy Meeting}\label{sec:NAM}

\subsection{Methodology}\label{sec:nammethod}

An informal survey of science fiction enthusiasm amongst professional astronomers was undertaken at the UK National Astronomy Meeting (NAM 2022) in July 2022. The National Astronomy Meeting is a relatively large conference, aimed at gathering together as many members as possible of the professional UK astronomy, solar and space physics communities, under the auspices of the Royal Astronomical Society. Historically, it has been focussed on relatively junior members of the community, with strong participation by PhD students and postdoctoral researchers, but with presentations from astronomers at all career stages. The 2022 meeting was hosted by the University of Warwick as a hybrid of both online and physical participation, and was the first in-person NAM (and for many the first in-person conference) for three years, following a cancelled meeting in 2020, and an online-only event in 2021 due to the pandemic.  It attracted 845 individual registrations, of which 643 attended in-person.

\begin{figure}
    \centering
    \includegraphics[width=\columnwidth]{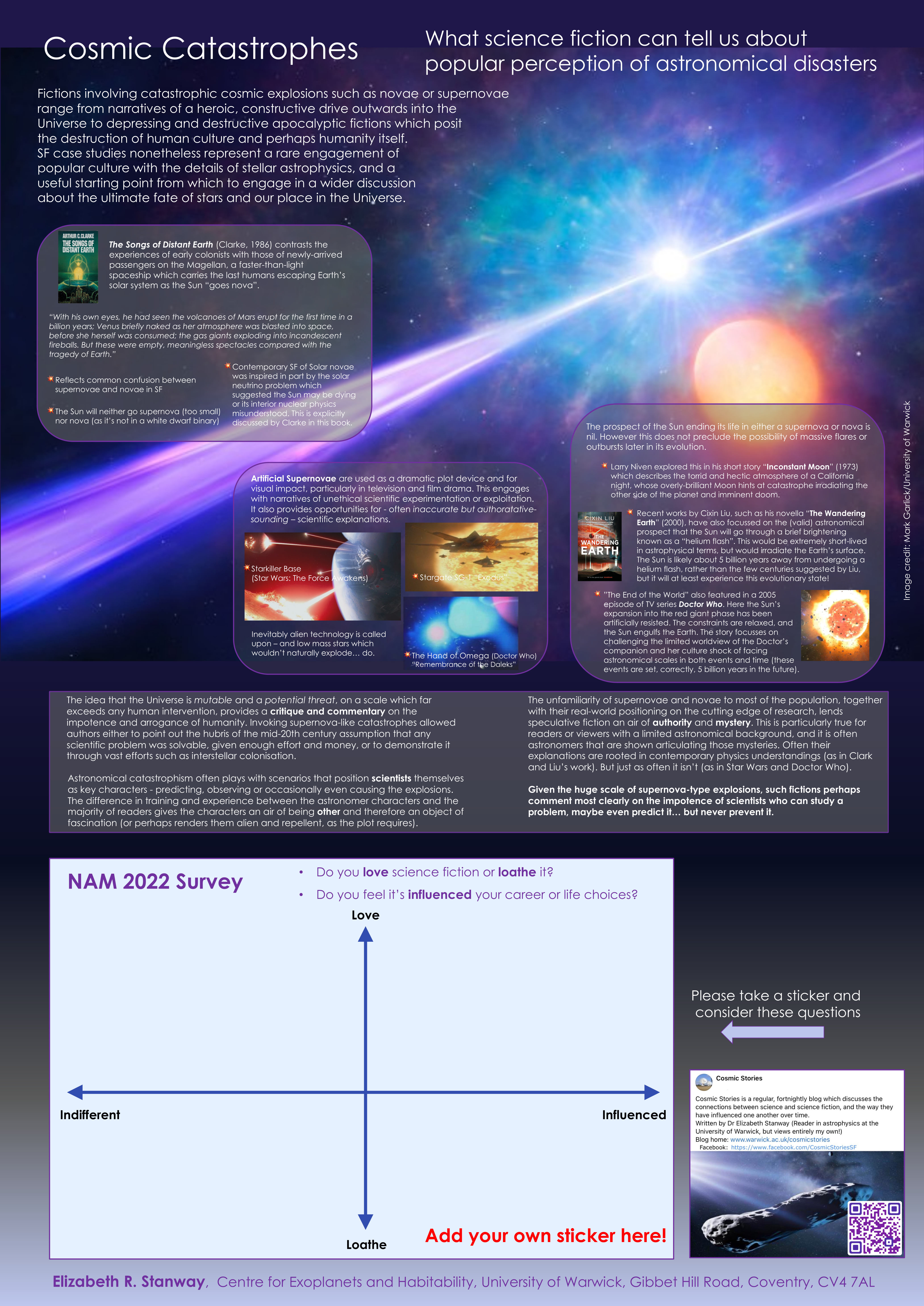}
    \caption{A0 (89$\times$119\,cm) survey poster presented at the UK National Astronomy Meeting in July 2022, showing the layout and context of the survey space. The poster was displayed (unsupervised) between Sunday evening and Friday afternoon in an exhibition hall. A sheet of small circular stickers was pinned in place towards the lower right to encourage responses posed to the questions in the lower third of the poster.}
    \label{fig:namposter}
\end{figure}

One element of the science programme was a display of approximately 180 A0-format (89$\times$119\,cm) scientific posters contributed by attendees. These were exhibited for the entire week, between a Sunday evening reception and Friday afternoon, in the same exhibition hall space used for serving refreshments and lunches. Together with a two hour poster session on Thursday afternoon, these periods encouraged attendees to circulate between and read the displayed posters. I prepared and presented a poster entitled “Cosmic Catastrophes: what science fiction can tell us about popular perceptions of astronomical disasters”. Two thirds of the poster presented the main content. The remaining third (approximately 42$\times$59\,cm)  was occupied by a box inviting audience participation in answering two questions:\\
\\
“Do you {\bf love} science fiction or {\bf loathe} it?”\\
“Do you feel it’s {\bf influenced} your career or life choices?”\\
\\
Responses were solicited in the form of small circular stickers to be placed in a two-dimensional parameter space defined by qualitative axes which varied from ‘loathe’ to ‘love’ in one direction and from ‘indifferent’ to ‘influenced’ in the other. These axes were isolated from the rest of the poster by a coloured background enclosing the survey within a box. The format and overall layout of the poster is shown in figure \ref{fig:namposter}. 

By construct, this was an entirely anonymous and unsupervised survey, providing a snapshot of the level of engagement in science fiction within the UK astronomical community\footnote{Note that theoretically the poster was also accessible to catering, cleaning and other conference venue support staff, although they were unlikely to interact with the conference content or contribute to this survey.}.

\subsection{Results}\label{sec:namresults}

\begin{figure}
    \centering
    \includegraphics[width=\columnwidth]{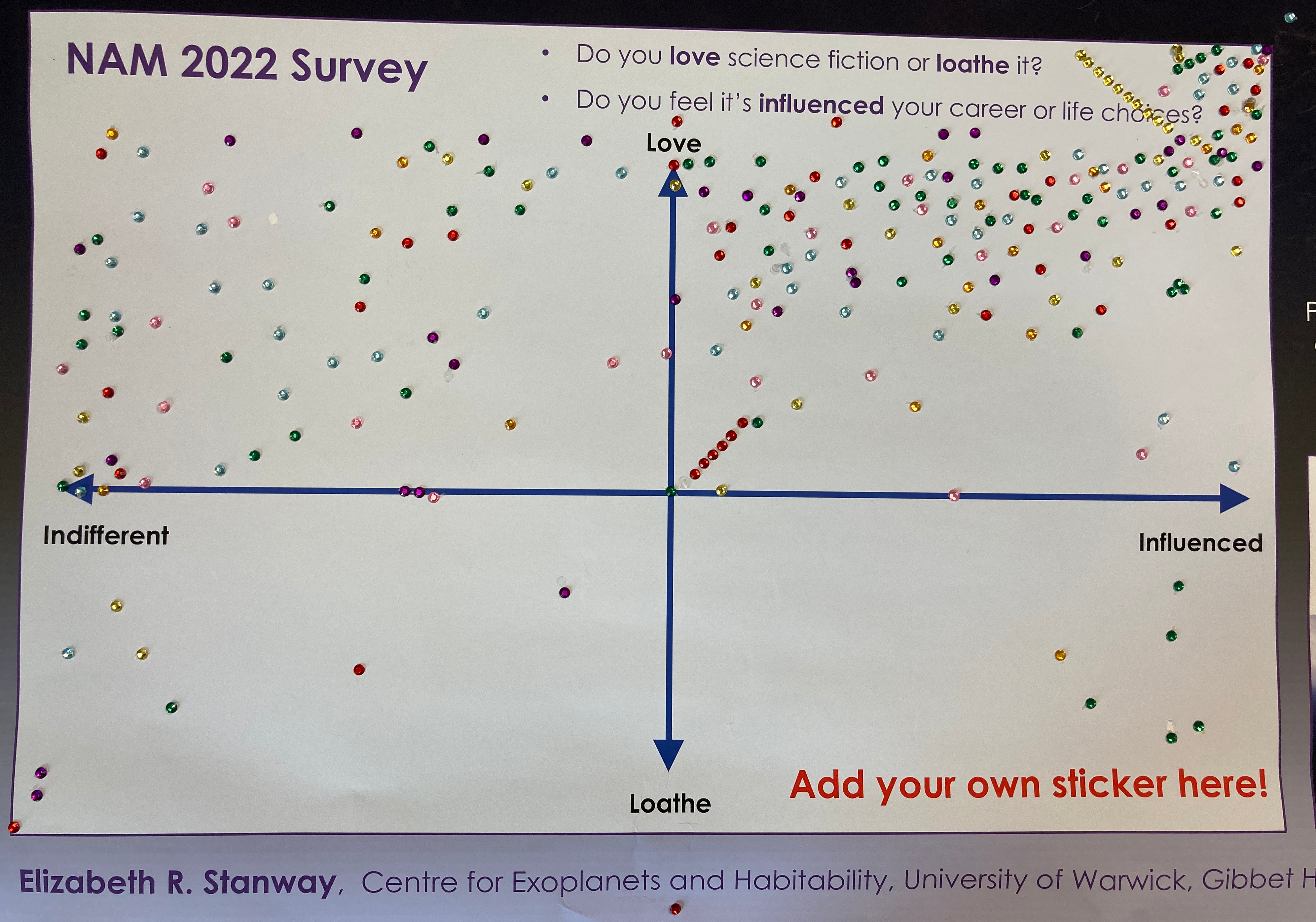}
    \caption{The final distribution of sticker-responses within (and adjacent to) the survey box, following NAM 2022.}
    \label{fig:posterresult}
\end{figure}

By the end of the meeting, 239 separate responses had been recorded (constituting more than a third of the 643 in-person attendees). The final distribution of responses in the survey box is shown in Figure \ref{fig:posterresult}. While the axes were deliberately qualitative, for this analysis responses have been digitised using a coordinate system where values of $\pm$1 indicates the extrema of the axes (with `love' and `influenced' assigned positive values). Digitised data is provided along with this manuscript. In two cases individual stickers were not separated from their neighbours on the provided sheet, and a string of connected markers was used. In these cases, only the least extreme value is counted as a response. 

\begin{figure}
    \centering
    \includegraphics[width=0.85\columnwidth]{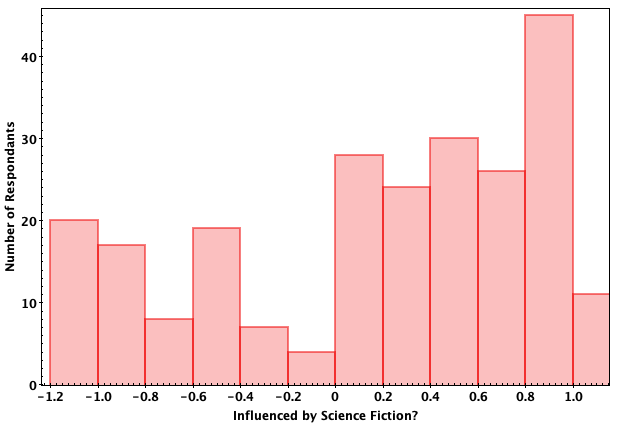}
    \includegraphics[width=0.85\columnwidth]{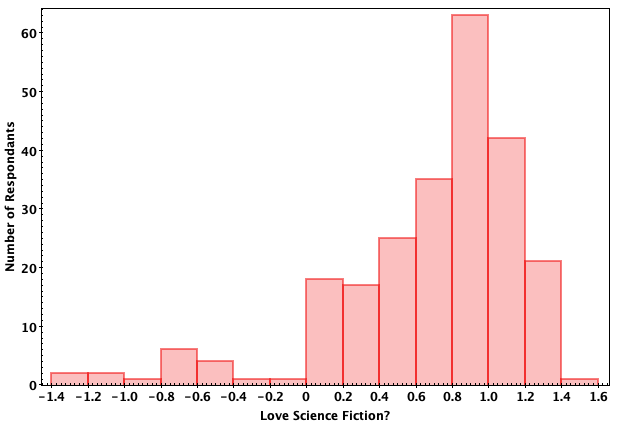}
    \caption{One-dimensional projections of the NAM 2022 survey responses, illustrating the distribution of astronomer responses to each question. In each case, values of $\pm$1 indicate the extreme of the axes in the survey space.}
    \label{fig:histograms}
\end{figure}

The distribution of points can be collapsed into one-dimensional histograms as shown in Figure \ref{fig:histograms}.  The distribution of astronomers is strongly peaked towards the extremum of the `love' axis. However the distribution of responses to the question of whether science fiction has been an influence is rather flatter, still favouring a positive effect, but with a substantial tail extending towards low values of influence.

A large majority of individual astronomers surveyed (66\%, 157/239) chose to place their response in the upper right quadrant of the survey (love/influenced), with many clustering around, or lying beyond, “love” as marked on the qualitative axis scale. A further 27\% of responses to the survey (66/239) occupied the upper left quadrant (love/indifferent).  Of 239 responses, only 16 (7\%) expressed significant dislike of science fiction - one of them going so far as to place a point at the extreme extension of the “loathe” axis possible while remaining on the poster. Ten further responses formally lay in the upper quadrants but were consistent with neutrality between like and dislike.  

\subsection{Interpretation}\label{sec:naminterp}

\begin{figure*}
    \centering
    \includegraphics[width=1.6\columnwidth]{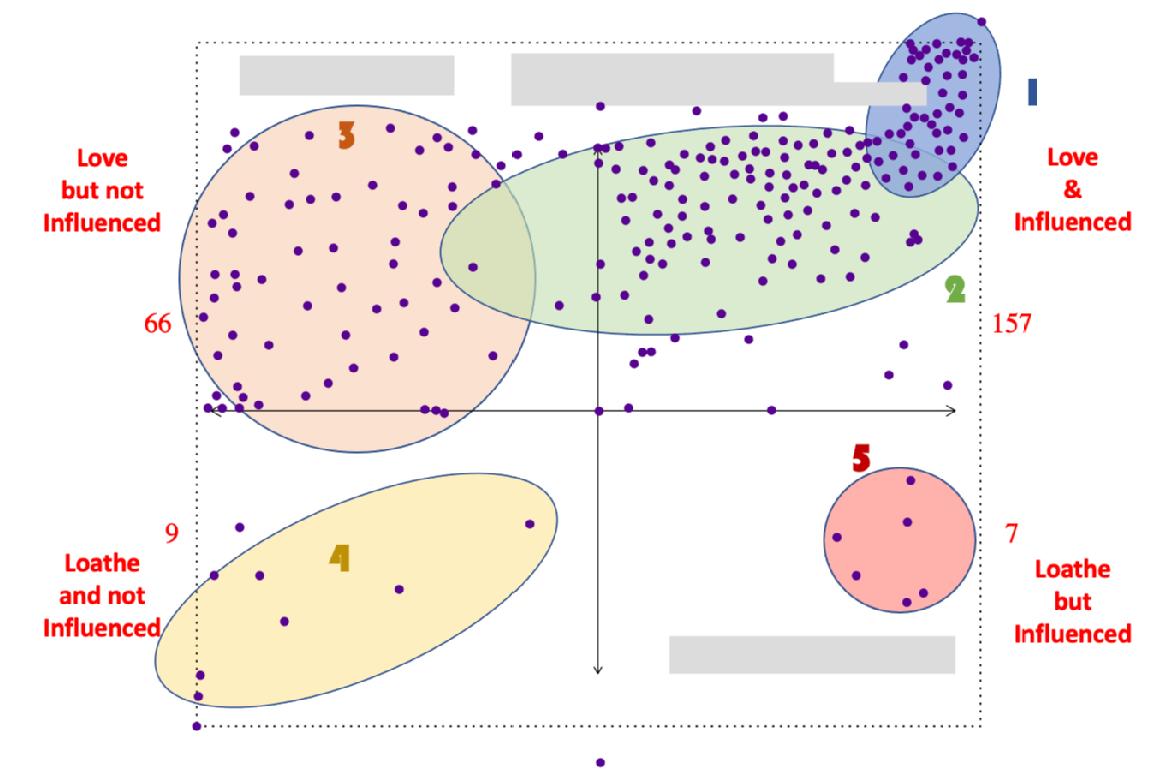}
    \caption{A schematic representation of the distribution of responses to the NAM 2022 survey, grouped into broad categories labelled 1 to 5, as explained in section \ref{sec:naminterp}. The dashed line indicates the limit of the survey box on the poster (note: two datapoints exceed this limit), while grey blocks indicate regions originally containing text. Small numbers outside the box indicate the number of data points in each quadrant.}
    \label{fig:schematic}
\end{figure*}

One possible schematic grouping of the responses received is shown in Figure \ref{fig:schematic}, with five broad regions indicated:

1) The asymptotic SF-lovers branch

2) The astronomers' main sequence

3) The weakly interacting cloud

4) The SFH (Science Fiction Haters) cooling track

5) The D clump


A clear majority of astronomers consider themselves both to love science fiction and to be influenced by it to a range of degrees, defining a \textit{main sequence}. At the extremum of this distribution, votes maximising both the love and influence expressed while remaining within the constraints of the survey box form an \textit{asymptotic branch}. A sizeable subset of astronomers are science fiction lovers, but do not identify as having been influenced by the genre (a \textit{weakly interacting cloud}), and a few outliers form interesting \textit{clumps}.

The existence of the asymptotic branch (reaching extreme values of “love” while at the limits of “influenced”) highlights the strength of feeling amongst those who are both passionate about science fiction and strongly influenced by it - many respondents made the effort to place their response in the most extreme position permitted by text and the survey box area. Given the size of this survey sample, there is clear evidence for inspiration from science fiction affecting the career choices of a significant fraction of professional astrophysicists.

The {weakly interacting cloud} constitutes around a quarter of responses to the survey. On average these are more likely to express a mild liking rather than strong passion for the subject. These, then, are the casual science fiction consumers, who do not connect their life choices with their viewing or reading preferences.

While the fraction of science-fiction dislikers is small, these will still constitute a few individuals in a typical astronomy group, who could easily feel alienated and isolated if a general liking for SF was presumed. Perhaps unsurprisingly the majority of these suggested that science fiction had not influenced them to any significant degree, with a weak trend towards diminishing influence with stronger loathing. However an interesting cluster of 6 individuals (the \textit{D clump}) identified as having been strongly influenced by science fiction while expressing a significant dislike for it. The motivation of these individuals is unclear. 
A possible interpretation might be that frustration with science fiction representations led such people to investigate science in reality, or that an initially positive SF influence has since turned into a strong dislike as they learnt more of the underlying physics. Further surveying would be required to investigate the career choices of this group further - although since they represent $<3$\% of the population, such a survey would necessarily be large or highly focussed.

\section{Discussion}\label{sec:disc}

Both surveys undertaken here provide statistical evidence supporting the hypothesis that professional astronomers are predominantly enthusiastic consumers of science fiction. The second survey, presented in section \ref{sec:NAM}, also provides strong evidence that science fiction is perceived to have had a powerful influence on the career choices of a substantial fraction of professional astronomers in the UK.

It is, of course, necessary to note limitations in the survey methodology. In neither survey reported here was demographic information regarding participants recorded, largely in the interests of simplifying the survey procedure and avoiding any perceived or psychological barrier to participation. There is no way of assessing the reasons for non-participation, particularly in the case of the second (NAM 2022) survey for which no narrative responses were recorded. While the first (Warwick) survey conflated enjoyment of science fiction with its influence, in the case of the NAM poster the survey axes presented were deliberately qualitative and invited free interpretation. As a result, some of the structure seen in the distribution of points was shaped by the layout of words and box shape on the poster and may not be intrinsic.
There was also no compelling reason for participants to take either survey seriously and give honest answers (although nor was there any obvious motivation for giving false answers).

The reasons for non-participation are perhaps the most concerning aspect of any statistical sampling analysis. If disengagement with the poster survey reflects disengagement with science fiction as a genre, then the statistics quoted here will be unrepresentative. In the case of the Warwick survey, participation was high (approximately 80\% of the active group at the time). Participation in the NAM poster was substantial but statistically more limited (37\% of potential in-person attendees).
However it is likely that many attendees remained unaware of the poster entirely. This is considered probable
given that: (i) this poster was only one of $\sim$180 present in the poster hall (with another 40 available online); (ii) a large but unknown fraction of attendees were present for only 1 or 2 days and prioritised science interaction over poster-reading in the time available; (iii) the conference schedule was crowded (with six science parallel sessions occurring in each timetable block, and typically two different lunch sessions available each day); and (iv) many conference attendees expressed concern regarding lingering in the crowded poster hall, given the ongoing pandemic. As a result, non-response is considered more likely to be motivated by a lack of awareness of the poster than an aversion to its content. If so the statistical breakdown of the community presented here is likely to be at least somewhat robust.

Despite the limitations noted above, the survey results  support the anecdotal evidence discussion in section \ref{sec:intro}, including the common use of science fiction references in astronomical software, notable examples of science fiction writers with professional astrophysics experience (including Patrick Moore, Fred Hoyle, Carl Sagan, Eric Kotani, Gregory Benford and Alistair Reynolds amongst others) and prominent science-fiction-fan space scientists. The NAM study further provides strong support for anecdotal evidence that science fiction can have a powerful motivating effect on the choice of astronomy careers. However it is important to acknowledge that there is, nonetheless, a substantial minority of astronomers for whom science fiction holds no interest, and who can feel alienated by an assumption to the contrary. The astrophysics community should be aware that science fiction references in papers, surveys, software or press engagement may be off-putting to, or actively disliked by, between about 10 and 20\% of their colleagues. 

Science fiction has been shown to have a complex impact on the public perception of science and scientists, particularly amongst children. While largely positive \citep[e.g.][]{laprise,Reinsborough2017-yv,doi:10.1177/2158244018780946,viggiano}, individual examples of particularly bad science fiction narratives have been identified as potentially misleading audiences regarding the underlying science \citep[e.g.][]{Barnett2006-tf}. Levels of science fiction enthusiasm amongst the wider public are substantially lower than those amongst professional astronomers: although robust statistics for the former are difficult to establish, science fiction films typically take a market share of between 10 and 20\% at the box office\footnote{\url{https://www.the-numbers.com/market/creative-type/Science-Fiction} [Accessed 3 Aug 2022].}. As a result, there is a risk that the astronomy community may alienate a fraction of their target audience amongst the general public while using SF examples and framing for public discourses regarding their results. 

Another concern may be presented by the culturally-influenced gender bias in science fiction engagement amongst the general population. Many recent science fiction television series and movies have been primarily targetted at, and thus responded to most strongly by, male audiences, with the substantive contributions to SF by female voices historically marginalised or overlooked. Future work may wish to address whether there is a similarly gendered division in the interaction between science fiction and career choice amongst astronomers  - an issue that has not been explored within the surveys reported here, but which has been identified as of potential concern in the use of science fiction for physics education \citep{2015CSSE...10..921H}. Indeed the stereotype of physical scientists as obsessive nerds, which has formed part of the popular imaginary for many years, associates an interest in science fiction with social awkwardness and a lack of contact with reality, often in young men. This stereotype, which has been prominent in mass media examples such as the television series {\em The Big Bang Theory} \citep[e.g.][]{1885-101514}, fails to respect the diversity of genders, experiences, interests, backgrounds and voices which contribute to the astronomical community. The use of SF for engagement may risk reinforcing this false stereotype unless appropriate examples are carefully selected. Such a risk has the potential to be easily overlooked in a community where SF enthusiasm is the norm.

As a result, this study highlights the need to carefully consider the audience for outreach or public engagement activities, particularly if other astronomers are used as test audiences or for feedback in the development process. A balance must be found between harnessing the inspiring qualities of science fiction narratives, particularly for aspiring astronomers, and effectively engaging a broader audience who may not themselves be enthusiastic consumers of the genre.

\section{Conclusions}\label{sec:conc}

In this report, I have presented the results of two surveys of science fiction engagement amongst the professional astrophysics community in the UK, with the following principle results:

\begin{enumerate}
    \item Amongst 36 members of the University of Warwick Astronomy \& Astrophysics research cluster, 80$\pm$14 percent of astronomers either liked or felt influenced by science fiction when surveyed in 2021. Strong interest was expressed in a range of science fiction franchises and authors. 
    \item A survey of 239 astronomers and space scientists undertaken at the UK National Astronomy Meeting 2022 indicates that 93$\pm$6 percent of UK astronomers liked or loved science fiction. 69$\pm$5 percent of astronomers believed that their life or career choices had been influenced by the genre, in many cases strongly so.
\end{enumerate}

To the best of my knowledge, there has not hitherto been a large, statistically robust survey to assess whether the established anecdotal examples of SF-engaged astronomers are indeed representative of the current professional astronomy community, or of the impact of SF on career selection in this field. The reasons for the trends reported here have not been probed in these studies, although it is likely that astronomy and science fiction appeal to the same people because of their underlying similarity in intent: to extend our understanding of the universe in which we live. 

Further work is required to more thoroughly probe the interconnection between science career choice and SF. However, as these surveys confirm, the visions of possibility in science fiction, the new horizons opened by speculative thought, and the potential impact of human imagination and creativity envisaged by science fiction writers, remain powerful motivators of scientists both of the present and of the future.

\acknowledgements

I would like to acknowledge the enthusiastic response of the astronomy community to my unofficial survey, and also thank Dr David Brown and the Local Organising Committee of NAM 2022 at the University of Warwick for organising the successful event. I also thank Dr Jan J. Eldridge for her support in this work and for assistance in naming features in response parameter space. 

\vspace{12pt}

\bibliographystyle{aasjournal}
\bibliography{ref}

\end{document}